\documentclass{aa}  

\usepackage{graphicx}
\usepackage{txfonts}
\usepackage{booktabs}
\usepackage{makecell}
\usepackage[dvipsnames]{xcolor}
\begin{document}

   \title{First hard X-ray imaging results by Solar Orbiter STIX}

   \subtitle{}

    \author{Paolo Massa\inst{1} \and 
          Andrea Francesco Battaglia\inst{2,3} \and 
          Anna Volpara \inst{1} \and 
           Hannah Collier\inst{2,3} \and 
           Gordon J. Hurford\inst{2} \and    
           Matej Kuhar\inst{2} \and  
            Emma Perracchione\inst{4} \and  
            Sara Garbarino\inst{1} \and
     	Anna Maria Massone\inst{1,10} \and    
      	Federico Benvenuto\inst{1} \and
       Frederic Schuller\inst{5} \and        
    	Alexander Warmuth\inst{5} \and        
     Ewan C. M. Dickson\inst{6} \and	
   Hualin Xiao\inst{2} \and
   Shane A. Maloney\inst{7,8} \and  	
   Daniel F. Ryan\inst{2} \and
   Michele Piana\inst{1,10} \and
      S\"am Krucker\inst{2,9}
     }

   \institute{
             MIDA, Dipartimento di Matematica, Università di Genova, via Dodecaneso 35, I-16146 Genova, Italy
             \and      
            University of Applied Sciences and Arts Northwestern Switzerland, Bahnhofstrasse 6, 5210 Windisch, Switzerland 
  	         \and
             ETH Z\"urich, R\"amistrasse 101, 8092 Z\"urich, Switzerland 
             \and
             Dipartimento di Matematica \lq\lq Giuseppe Peano\rq\rq, Università di Torino, Via Carlo Alberto 10, 10123 Torino, Italy
 		     \and 
              Leibniz-Institut f\"ur Astrophysik Potsdam (AIP), An der Sternwarte 16, D-14482 Potsdam, Germany
            \and
	         Institute of Physics, University of Graz, A-8010 Graz, Austria
	         \and
 	         Astrophysics Research Group, School of Physics, Trinity College Dublin, Dublin 2, Ireland
 	         \and 
             School of Cosmic Physics, Dublin Institute for Advanced Studies, 31 Fitzwilliam Place, Dublin, D02 XF86, Ireland
 	         \and
             Space Sciences Laboratory, University of California, 7 Gauss Way, 94720 Berkeley, USA
             \and
             CNR - SPIN Genova, via Dodecaneso 33 16146 Genova, Italy
}
   \date{\today}

 
  \abstract
  {The Spectrometer/Telescope for Imaging X-rays (STIX) is one of 6 remote sensing instruments on--board Solar Orbiter. It provides hard X-ray imaging spectroscopy of solar flares by sampling the Fourier transform of the incoming flux. }
  {To show that the visibility amplitude and phase calibration of 24 out of 30 STIX sub--collimators is well advanced and that a set of imaging methods is able to provide the first hard X-ray images of the flaring Sun from Solar Orbiter.}
  {We applied four visibility--based image reconstruction methods and a count--based one to calibrated STIX observations.
  The resulting reconstructions are compared to those provided by an optimization algorithm used for fitting the amplitudes of STIX visibilities.}
  {When applied to six flares with GOES class between C4 and M4 which occurred in May 2021, the five imaging methods produce results morphologically consistent with the ones provided by the Atmospheric Imaging Assembly on-board the Solar Dynamic Observatory (SDO/AIA) in UV wavelengths. The $\chi^2$ values and the parameters of the reconstructed sources are comparable between methods, thus confirming their robustness.}
  {This paper shows that the current calibration of the main part of STIX sub--collimators has reached a satisfactory level for scientific data exploitation, and that the imaging algorithms already available in the STIX data analysis software provide reliable and robust reconstructions of the morphology of solar flares.}



   \maketitle
%
\section{Introduction}

The Spectrometer/Telescope for Imaging X-rays (STIX) on Solar Orbiter studies solar flares in hard X-ray wavelengths \citep{2020A&A...642A..15K}. STIX imaging is not based on focusing optics; instead, it exploits 
an
imaging technique realized by means of $30$ pairs of Tungsten grids
\citep{hurford2013x} placed in front of coarsely pixelated CdTe detectors \citep{2015NIMPA.787...72M}.
As a consequence, STIX is a Fourier--based imager that provides $30$ samples of the Fourier transform of the incoming photon flux, termed \emph{visibilities}, and STIX images at different photon energies can be reconstructed by means of algorithms for the inversion of the Fourier transform from limited data \citep{pianabook,2021InvPr..37j5001P}. A first version of the calibration of STIX visibility amplitudes became available in Spring 2021, and a demonstration of STIX imaging capabilities using these semi-calibrated visibilities has been obtained by using parametric approaches \citep{massa2021imaging}. 
Since then, the calibration of both amplitude and phase of STIX visibilities has been carried out for the 24 coarsest sub--collimators labeled from 3a through 10c, where the number refers to the detector resolution and the letter a, b, or c refers to the orientation of the grids \citep[see Table 2 in][]{2020A&A...642A..15K}.
Sub--collimators 1 and 2 have the finest angular resolution and were fabricated differently than the coarsest ones.
The complete calibration for these is in progress.

The objective of this paper is two--fold. First, this paper demonstrates that the complete calibration of 24 out of 30 STIX visibilities has reached a satisfactory level for accurate image reconstruction; second, this work demonstrates that several image reconstruction methods can produce comparable, robust and reliable results. 
The individual steps involved in the STIX imaging calibration specific for this instrument, and they will be published in a future paper where all the details will be discussed. Here we concentrate on the first on the first reconstructions from calibrated data. Compared to the phase calibration in radio astronomy where the calibration changes in time with changing atmospheric conditions, the STIX phase calibration depends only on the mechanical properties of the grids relative to the detectors and the calibration is therefore stable in time. This enormously simplifies the calibration task for STIX, as a single calibration is good for all flares.

Our analysis is focused on a set of flaring events that occurred in May 2021. These flares range from GOES C4 to M4 class, and all of them have good counting statistics in the STIX observations. 
Further, the May 7 event can be considered as paradigmatic of the standard flare model with two non--thermal footpoints connected by a flare loop \citep{1988psf..book.....T}. This event is therefore particularly significant for the validation of the calibration process and of the imaging algorithms' performances.

As there is currently no other solar--dedicated hard X-ray imaging telescope observing the Sun, we have no means of comparing STIX observations with other hard X-ray images. For assessing the reliability of the reconstructed flare morphology, the STIX images are compared to Ultraviolet (UV) maps of the same events provided by the Atmospheric Imaging Assembly on-board the Solar Dynamic Observatory (SDO/AIA) \citep{2012SoPh..275...17L}. To account for the different vantage points of the SDO and Solar Orbiter, the AIA UV images are rotated to the STIX reference frame
under the assumption that the emission comes from the solar surface. Such a rotation is accurate enough for our purpose (i.e. a few arcsec) as most of the UV emission originates from the chromosphere \citep[e.g.,][]{2004SoPh..222..279F}. 

The plan of the paper is as follows. Section 2 details the STIX observations utilized for the experiments and provides a brief overview of the imaging methods. Section 3 contains the results of this study. Our conclusions are offered in Section 4.

\section{STIX images of May 2021 events}

\begin{figure*}[tbp]
\centering 
\includegraphics[width=0.98\textwidth,keepaspectratio,trim=6 285 35 0, clip]{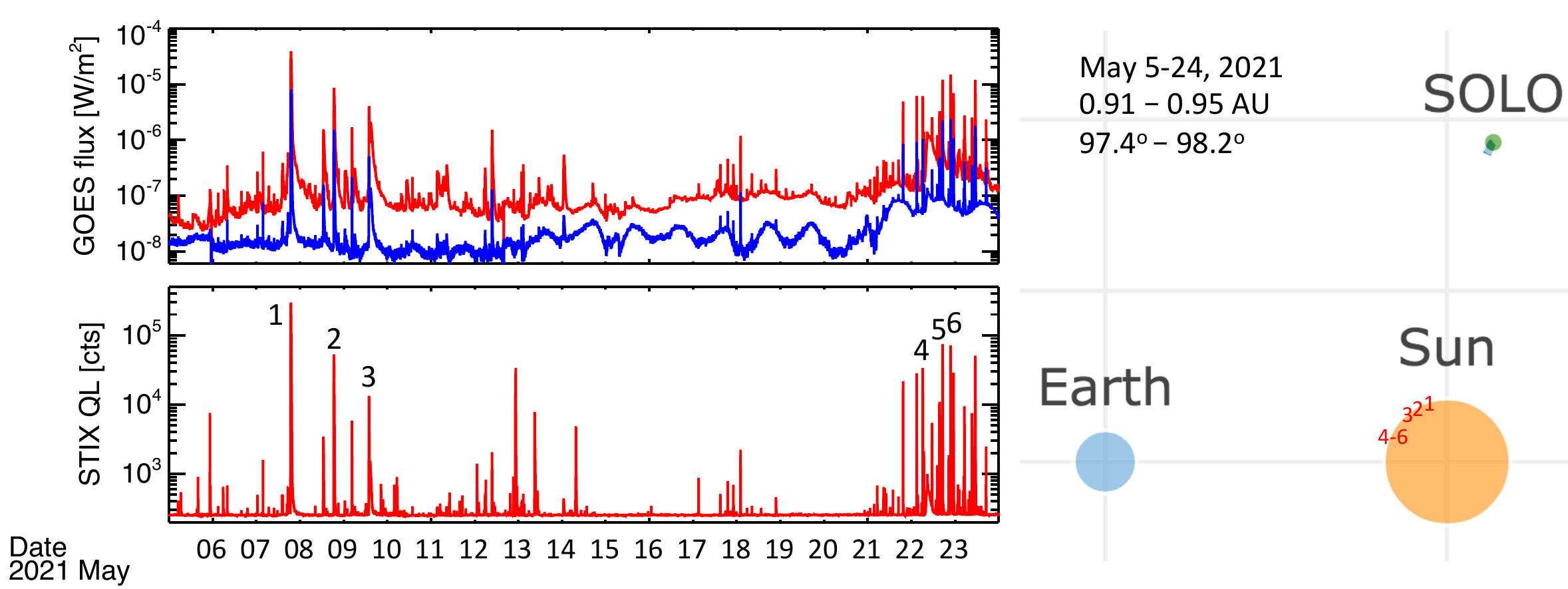}
\caption{GOES Soft X-ray profiles (\emph{top}; low energy channel in red, high energy channel in blue) and STIX 4-10 keV quicklook lightcurve (\emph{bottom}) during the time period considered in the paper. The six flares selected for the STIX imaging studies are marked by numbers. On the right part of the figure the relative position of Solar Orbiter, Earth, and Sun are shown for the same period, as well as the locations of the six flares on the Sun. Note that the relative position is only slightly changing as Solar Orbiter and Earth orbital velocities are rather similar during the month of May 2021.}
\label{Fig:qllc}
\end{figure*}

\subsection{Data overview}

In May 2021 two active regions (ARs) generated a series of C and M class flares observed on disk by Solar Orbiter STIX as well as by GOES and SDO/AIA.
The GOES and STIX lightcurves of all events from May 5 to May 24 are given in the top and bottom left panels of Figure \ref{Fig:qllc}.
To demonstrate the imaging capability of STIX, we focused on imaging during the impulsive phase of six flares: three flares associated to AR2822 which occurred on May 7 (GOES M3.9; 18:51:00--18:53:40 UT), May 8 (GOES C8.6; 18:29:00--18:32:00 UT), and May 9 (GOES C4.0; 13:53:00--13:55:00 UT); and three flares from AR2824 which all occurred on May 22 in the time ranges 02:52:00--02:55:00 UT (GOES C6.1), 17:08:00--17:11:00 UT (GOES M1.1), and 21:30:30--21:34:00 UT (GOES M1.4). 
We point out that the visibility data associated with these events have been calibrated 
using parameters and procedures as of January 2022.
It is important to highlight that second order corrections to the calibrations are still in progress and promise to improve the imaging quality in future.


\subsection{Spectroscopy}

\begin{figure*}
    \centering
    \includegraphics[width=\textwidth]{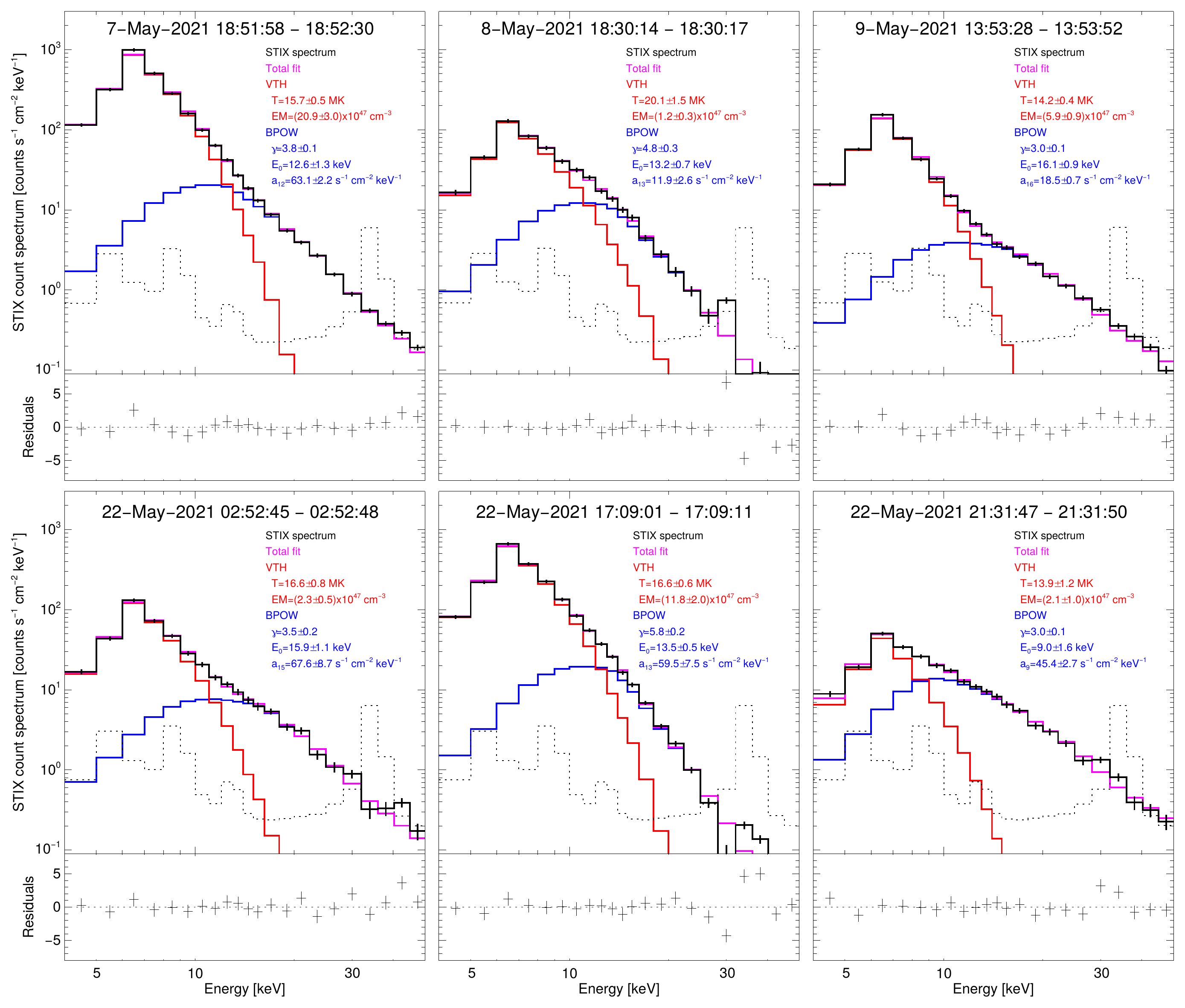}
    \caption{Spectroscopic results of the six events considered in this paper. For each event, the background subtracted count spectrum, given in black, is obtained by integrating the STIX temporal profiles around the non--thermal peak time. The STIX background spectrum used for the subtraction is shown in dotted black. The colored curves represent the fitted components: single thermal \texttt{‘vth’} in red and non--thermal broken power-law \texttt{‘bpow’} in blue, while the magenta curve denotes their sum. The bottom panel of each spectrum shows the residuals, i.e. the difference between the observed STIX count spectrum and the total fit normalized by the errors derived from counting statistics.  
    }
    \label{fig:spectra}
\end{figure*}

The spectroscopic analysis of the six events considered in this paper is done using the OSPEX\footnote{\url{http://hesperia.gsfc.nasa.gov/ssw/packages/spex/doc/}} SSWIDL software (release from September 2021) and is presented in Figure~\ref{fig:spectra}.
The main goal of the spectral fitting analysis done in this paper is to find suitable energy ranges to use for imaging of the thermal and the non--thermal emission. While the selection of the thermal energy range is rather straightforward (i.e. around the peak in the count spectrum at 6-7 keV), it is desirable to extend the non-thermal range to as low energies as possible to enhance statistics in the derived visibilities, but care should be taken to avoid any contribution of thermal emissions.
To include thermal emissions would make the image morphology more complex.
For example, in the standard flare model three sources would exist in this case, instead of just two footpoints, which would significantly increase the difficulties of image reconstruction with our limited sample of visibilities.

The spectral plots in Figure 2 are scaled in the same way in energy and rate making it straightforward to compare the different events (for information on STIX spectral fitting we refer to \citet{2021A&A...656A...4B}). As the time intervals are selected at the peak time of the non--thermal emission, which is generally earlier than the GOES peak time, we cannot compare directly the STIX thermal fits with GOES class of these events. The non--thermal emission, however, can be directly compared revealing a significant spread between events, typical for solar flares \citep[e.g.,][]{2005A&A...439..737B}. 
The power law indices of the photon spectra show rather hard spectra for 4 flares with values below $\gamma$=4, while the event on May 22 around 17UT has a very soft spectrum with a $\gamma$ around 6, a rather high value considering that it is a M1 class flare.  As most often is the case in X-ray spectral analysis of solar flares, the fitted energy break is driven by the existence of thermal emission at lower energies and it does not necessarily  indicate the existence of turnover in the distribution of the accelerated electrons below the fitted break energy.  However, we used these values to make sure that the energy range for imaging the non--thermal footpoints starts at a higher value than the fitted break energy.  

For all of these medium large flares, the calibration line of the Ba133 source outshines the flare counts at the energy bin that contains the 31 keV line.  
As the calibration source is essentially stable in time during a flare, background subtraction often works well, even when the flare signal is $\sim$10 times lower than the background.  For energy bins above the calibration line, the flare is roughly at the same level as the background emission, at least for the events shown here.
Due to a strong calibration line in the 28-32 keV energy bin, image reconstruction including this energy bin can be challenging.
Indeed, for flares that are not strong enough, the error on the background emission can be larger than the flaring signal itself. In this case, background subtraction can remove any information on the flaring source contained in the count measurements (and hence, in the visibilities).
We are currently further investigating this point. 
As a current best practice, we recommend to avoid making an image with an energy range starting at 28 keV, as in such a case the background likely dominates the signal. 
Similarly, it does not make much sense to include the 28-32 keV channel as the last energy bin in an imaging energy range, as the proportional increase of the background is most often larger than the increase in signal. Hence, for all but one flare in this paper the highest energy bin included for imaging is 28 keV. To include the 28-32 keV bin as an intermediate bin in the selected imaging energy range made no  significant difference, for at least the May 7 event presented here (the image presented for the flare of May 7 has the 28-32 keV bin included in the range 22 to 50 keV). In any case, the effect of enhanced background emission due to the calibration line at 31 keV can be investigated when an image which includes the 28-32keV energy bin is compared with the corresponding image excluding that energy bin.

\subsection{STIX and AIA}

For validating the morphology of the reconstructed hard X-ray sources, we performed a comparison with the 1600 \AA{} images provided by SDO/AIA for all six events. To do so, we needed to account for the relative position of STIX and AIA (Earth) with respect to the location of the events on the Sun (Figure \ref{Fig:qllc}, right panel).
The AIA images have been rotated to the Solar Orbiter vantage point by means of the \textit{reproject}\footnote{\url{https://reproject.readthedocs.io/en/stable/}} Python package, where the Solar Orbiter ephemeris have been obtained from the Operational Solar Orbiter SPICE Kernel Dataset\footnote{\url{https://doi.org/10.5270/esa-kt1577e}}, following the procedure used in \citet{2021A&A...656A...4B}.
A potentially significant source of error on the location of the flare emissions in the rotated AIA maps is given by the projection effects. Indeed, coronal structures that extend higher up in the atmosphere will be distorted due to projection effects once the rotation of the map is performed. In order to minimize such uncertainty, one has to apply this method only to maps in which the observed emission originates roughly from the same altitudes, and ideally close to the solar surface. This can be the case for the emission that is observed in flare ribbons in the AIA 1600 and 1700 \AA{} maps. In this paper, we compare the STIX reconstructed images with the rotated AIA 1600 \AA{} maps as they would be seen from the Solar Orbiter vantage point. However, a word of caution is required here. 
The routine for rotating the AIA maps assumes all the emission is coming from the solar surface, i.e., from the bottom of the photosphere. That is not exactly what the AIA 1600 \AA{} shows, since, according to the standard flare scenario, the emission mostly comes from the chromosphere. This may eventually result in a systematic offset in the location of flare ribbons as seen from the Solar Orbiter vantage point, which also depends on the location of Solar Orbiter relative to Earth and its distance from the Sun. In the case of the flares considered in this paper, we estimated this uncertainty to be roughly 0.9 arcsec for the May 7, May 8 and May 9 events and 1.4 arcsec for the May 22 events, which is much smaller than the angular resolution of the finest sub--collimators used at 14.6 arcsec. We therefore neglected this second order effect.

\subsection{Location of the STIX reconstructions}

The STIX reconstructions are performed in an instrument reference frame. 
In order to plot them in the helioprojective coordinate system, we need to perform a roto--translation that takes into account the spacecraft roll--angle and the angular offset between the instrument optical axis and spacecraft reference axis.
The roll-angle value used for the rotation is provided by the SPICE as-flown attitude kernel\footnote{\url{https://doi.org/10.5270/esa-kt1577e}}.
As for the offset issue issue, for the events considered in this study, Solar Orbiter was too far (outside 0.75 AU) from the Sun for the STIX Aspect System to be able to provide absolute pointing information \citep{warmuth2020stix}.
In order to superimpose the reconstructed hard X-ray sources on the UV ribbons, we therefore performed a manual shift of the reconstructed and rotated STIX maps. 
The shift, however, should not be arbitrarily large. 
For times when the aspect solution is available and the spacecraft is pointing at solar center (except for campaigns, Solar Orbiter is most of the time pointing at solar center), the absolute pointing of the STIX images without applying the aspect correction should be less than 100$''$ offset from the actual value. Furthermore, the shift should not be drastically different for individual flares.
The values we used for the six flares presented here are given in Table \ref{tab:shifts} and they indicate that the applied shifts are reasonable.

\begin{table}[ht]
\centering
\begin{tabular}{ccc}
\toprule
Event                   &$\Delta x$ (arcsec)    &$\Delta y$ (arcsec)\\
\midrule
May 7 2021 18:51:00     &$44$                   &$54$\\
May 8 2021 18:29:00     &$45$                   &$57$\\
May 9 2021 13:53:00     &$47.5$                 &$53$\\
May 22 2021 02:52:00    &$47$                   &$55$\\
May 22 2021 17:08:00    &$47$                   &$50$\\
May 22 2021 21:30:30    &$50$                   &$50$\\
\bottomrule
\end{tabular}
\caption{Shift applied to the $x$ and $y$ coordinates of the center of the STIX reconstructions shown in Figures \ref{fig:active-region-1} and \ref{fig:active-region-2} to make them overlap with the flare ribbons in the AIA maps.
The $x$ and $y$ shifts are defined in the heliocentric coordinate system and increase towards solar West and solar North, respectively.}
\label{tab:shifts}
\end{table}

We point out that the image placement accuracy issue will not represent a problem during 
the Solar Orbiter mission science phase,
which started at the end of November 2021. During the official science window, Solar Orbiter will be close enough to the Sun to allow precise measurements of STIX pointing by means of the Aspect System. 
We expect the error in the location of STIX reconstructions to be better than 4 arcsec \citep{warmuth2020stix} once the corrections based on these measurements are implemented.

\subsection{Imaging methods}

\begin{table*}[ht]
\centering
\resizebox{\linewidth}{!}{\begin{tabular}{rcccc}

\toprule

Method      & Input data    &\multicolumn{3}{c}{Parameters (value used in the paper)}\\
\midrule

Back Projection & Visibilities  &\makecell{\textbf{Weighting} \\ (natural)}    &   &\\

\midrule

Clean      & Visibilities   &\makecell{\textbf{Weighting Back Projection} \\ (natural)} &\makecell{\textbf{Beam width - FWHM} \\ (16.5 arcsec)}    &\makecell{\textbf{Clean boxes} \\ (50\% contour level)}\\

\midrule

MEM\_GE     & Visibilities    &\makecell{\textbf{Regularization parameter} \\ (0.005 for thermal, \\ 0.02 for non--thermal)} &\makecell{\textbf{Total flux estimate} \\ (max visib. amplitudes)} &\\

\midrule

EM      & Counts     &\makecell{\textbf{Stopping rule tolerance} \\ ($10^{-4}$)}  &   &\\

\midrule

VIS\_FWDFIT     & Visibilities    &\makecell{\textbf{Source shape} \\ (Gaussian elliptical source for thermal, \\ double Gaussian circular source \\ for non--thermal)}      &   &\\

\midrule

Amplitude fitting PSO   & Visibility amplitudes    &\makecell{\textbf{Source shape} \\ (Gaussian elliptical source for thermal, \\ double Gaussian circular source \\ for non--thermal)}    &\makecell{\textbf{Parameter uncertainty} \\ (on)}\\
\bottomrule
\end{tabular}}
\caption{Parameters and input data of the algorithms used for solving the STIX image reconstruction problem. 
For each parameter, the value set for obtaining the reconstructions presented in the paper is reported between brackets.}
\label{tab:param-algo}
\end{table*}

A set of image reconstruction methods is already implemented in the STIX data analysis software which will be made available in February 2022. Specifically, here we considered:
\begin{itemize}
    \item the Maximum Entropy Method MEM$\_$GE \citep{massa2020mem_ge}, which realizes the $\chi^2$ minimization with respect to the observed visibilities, under three constraints: maximum entropy, positivity of the reconstructed signal, and a flux constraint.
    Specifically, the latter forces the sum of the image pixels to be equal to an a priori estimate of the emitted total flux. 
    This estimate is obtained from the visibility values, hence the flux constraint does not add any further information on the solution.
    However, this constraint is needed for simplifying the optimization problem.
    MEM$\_$GE represents a computational upgrade of the MEM$\_$NJIT algorithm implemented for the Reuven Ramaty High Energy Solar Spectroscopic Imager (RHESSI) \citep{2006ApJ...636.1159B, Schmahl}, since it involves a mathematically sound optimization problem and a robust optimization technique.
    \item Back projection  \citep[see e.g.,][]{2002SoPh}, which realizes the discrete Fourier transform inversion of STIX visibilities and is equivalent to the \emph{dirty map} in radio interferometry.
    This Fourier integration can be computed using different quadrature formulae.
    Although arbitrary weights can be used, the two main choices utilize the same weighting for all visibilities (\emph{natural weighting}) or utilize a weighting that accounts for the distribution of visibilities in the frequency plane (\emph{uniform weighting}).
    The former weighting has been applied in this paper.
    
    \item Clean  \citep[see e.g.,][]{1974A&AS...15..417H}, which provides a set of point sources (named Clean Components) by iteratively deconvolving the instrumental point spread function (PSF) from a dirty map, which is provided by Back Projection with natural weighting in this case.
    While the Clean Components are the actual result, the Clean process is extended by two standard steps for visualization purposes: the convolution of the Clean Components with an idealized PSF and the addition of the residual map.
    
    \item VIS$\_$FWDFIT  \citep[see e.g.,][]{2002SoPh}, which minimizes the difference between the measured visibilities and those predicted by assumed sources (single and double Gaussian circular source, Gaussian elliptical source, loop). The algorithm utilizes the AMOEBA function to perform minimization \citep{press2007numerical}.
    An estimate of the retrieved parameter uncertainty is obtained by perturbing several times the data with Gaussian noise, by forward--fitting the perturbed data, and by computing the standard deviation of the set of optimized parameters.
    \item EM \citep{2019A&A...624A.130M}, which is the Expectation Maximization algorithm, also known as the Richardson-Lucy algorithm when applied to image deconvolution problems \citep{richardson1972bayesian, lucy1974iterative}. 
    EM takes as input the measured STIX counts instead of the corresponding visibilities. 
    The calibration of the counts follows the count formation model described in \citet{2019A&A...624A.130M} and exploits the same correction factors introduced for the visibility calibration. 
    Following a maximum--likelihood approach, EM starts from a constant image and uses an iterative scheme based on the discrepancy between the observed counts and those predicted from the current iterate through the forward count formation model. The algorithm finds the image which maximizes the probability that the observed counts are a realization of the Poisson random variable whose mean value is the array of the predicted counts.
    A positivity constraint is also imposed on the solution.

\end{itemize}
The results provided by these five approaches have been compared with the ones provided by a visibility amplitude fitting method based on Particle Swarm Optimization (PSO). 
Amplitude fitting, which has been validated in a previously published paper by \citet{massa2021imaging}, was an intermediate approach to extract imaging information from STIX observations before the phase calibration reached a satisfactory level. With the phase calibration now available, this method is essentially obsolete. Nevertheless, it is worth to compare this algorithm with the imaging methods presented in this paper.
Amplitude fitting realizes forward--fit by means of an optimization scheme simulating social behavior, specifically swarm intelligence \citep{clerc2010particle}.
The method also provides an estimate of the uncertainty on the retrieved parameters in a way analogous to VIS\_FWDFIT.
Table \ref{tab:param-algo} summarizes the main parameters of the algorithms and reports the values we set for performing the reconstructions.
The STIX data (visibilities, counts or visibility amplitudes) taken as input by each method are also indicated.

\section{Results}

\begin{figure*}[t]
\centering
\includegraphics[height=\textwidth,angle=90]{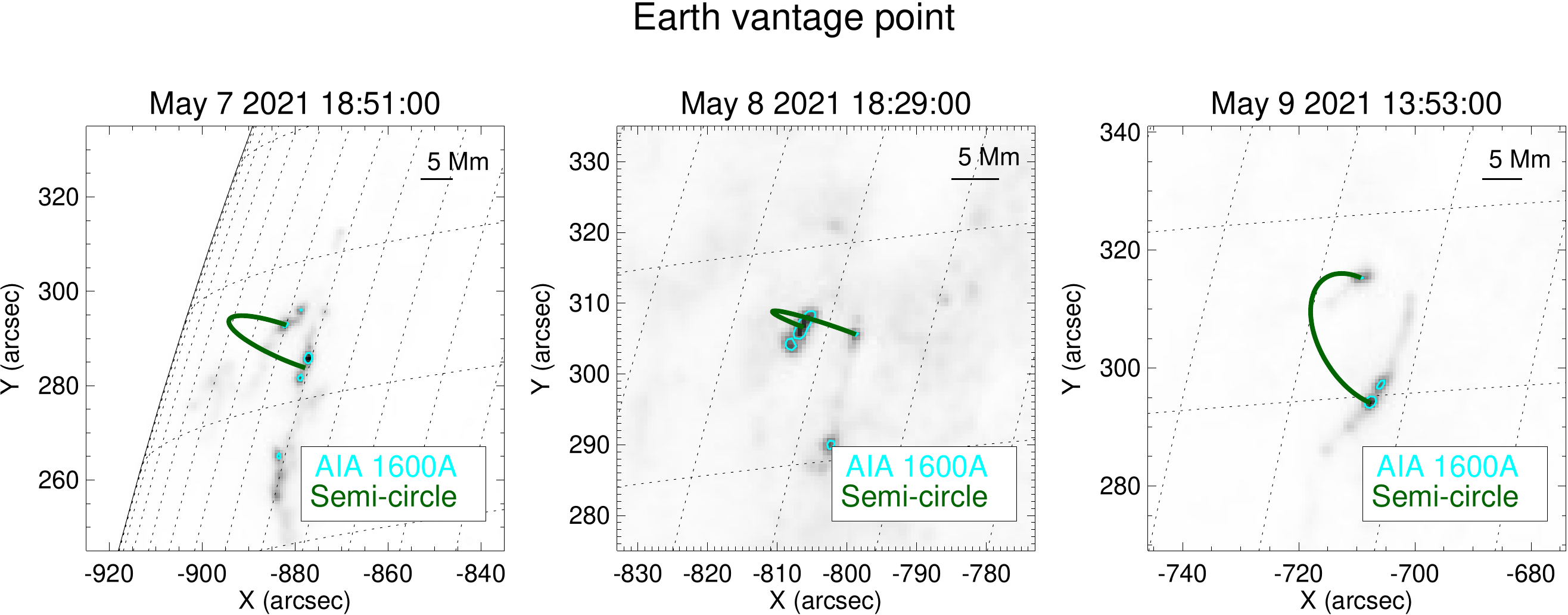}\\
\vspace{5pt}
\includegraphics[height=\textwidth,angle=90]{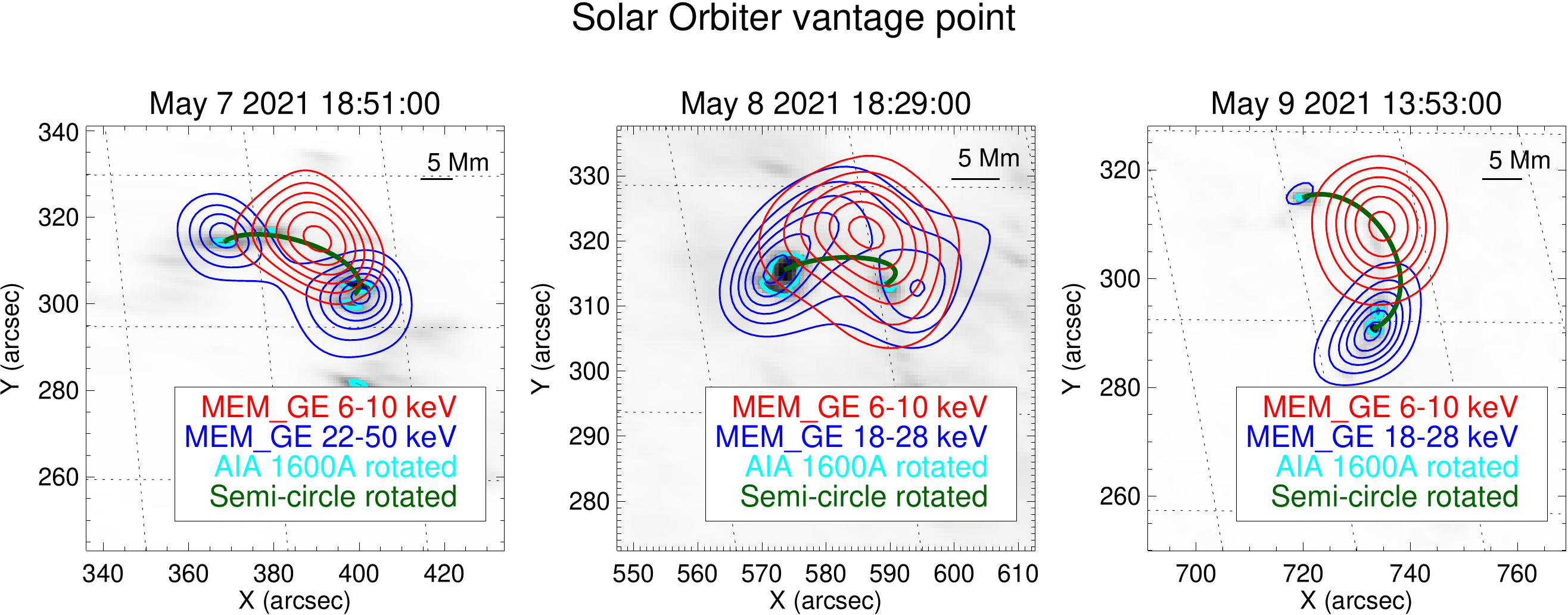}\\
\caption{Top row, from left to right: AIA 1600\,\AA{} images of the May 7 2021, May 8 2021 and May 9 2021 events, respectively. 
Bottom row: MEM$\_$GE reconstructions overlaid on the rotated AIA maps of the same events. The 50\% contour levels of the AIA images are plotted in cyan, while the 20, 35, 50, 65, 80 and 95\% contour levels of the reconstructed thermal and non--thermal X-ray emissions are plotted in red and in blue, respectively.
The energy intervals considered are 6-10 keV for the thermal components, 22-50 keV for the non--thermal component of the May 7 event and 18-28 keV for the non--thermal component of the remaining events.
As a reference, a semi--circle connecting the flare ribbons is plotted in dark green as seen from the Earth and as seen from Solar Orbiter in the top--row and in the bottom--row panels, respectively.
A horizontal bar indicating a length of 5 Mm on the plane of the sky is reported in the top--right corner of each plot.}
\label{fig:active-region-1}
\end{figure*}

\begin{figure*}[t]
\centering
\includegraphics[height=\textwidth,angle=90]{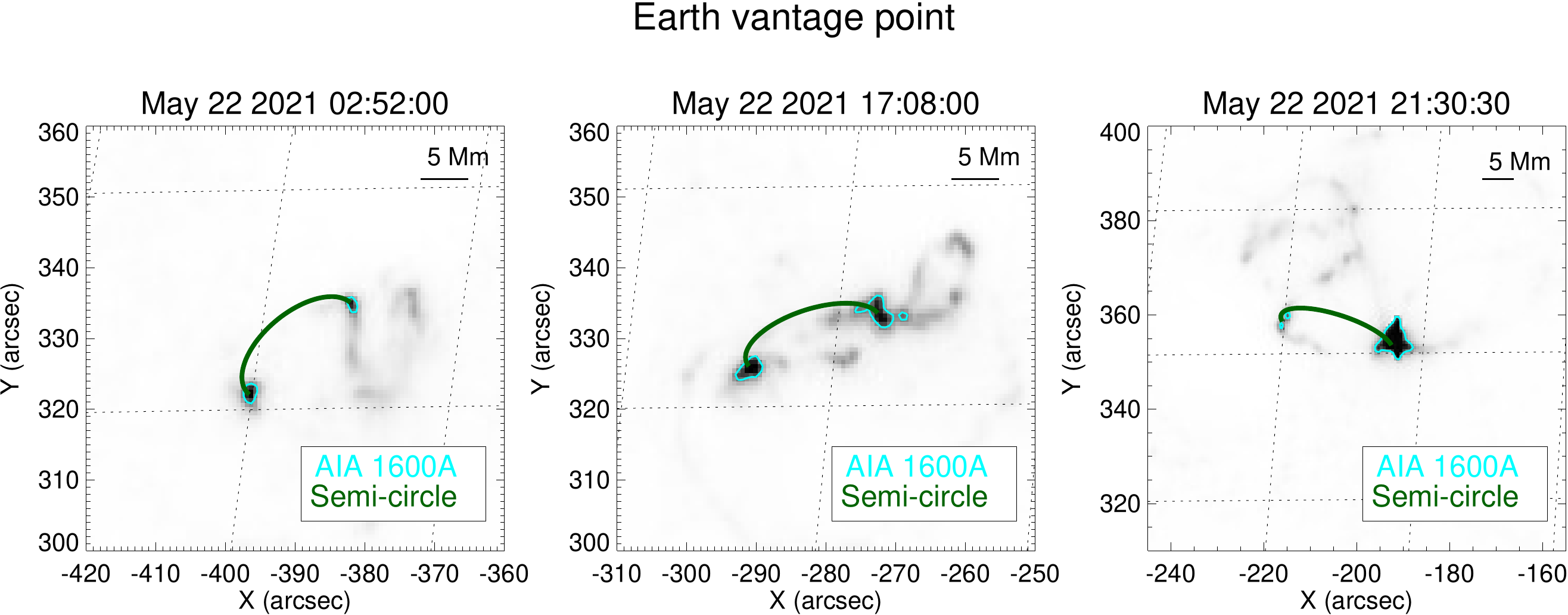}\\
\vspace{10pt}
\includegraphics[height=\textwidth, angle=90]{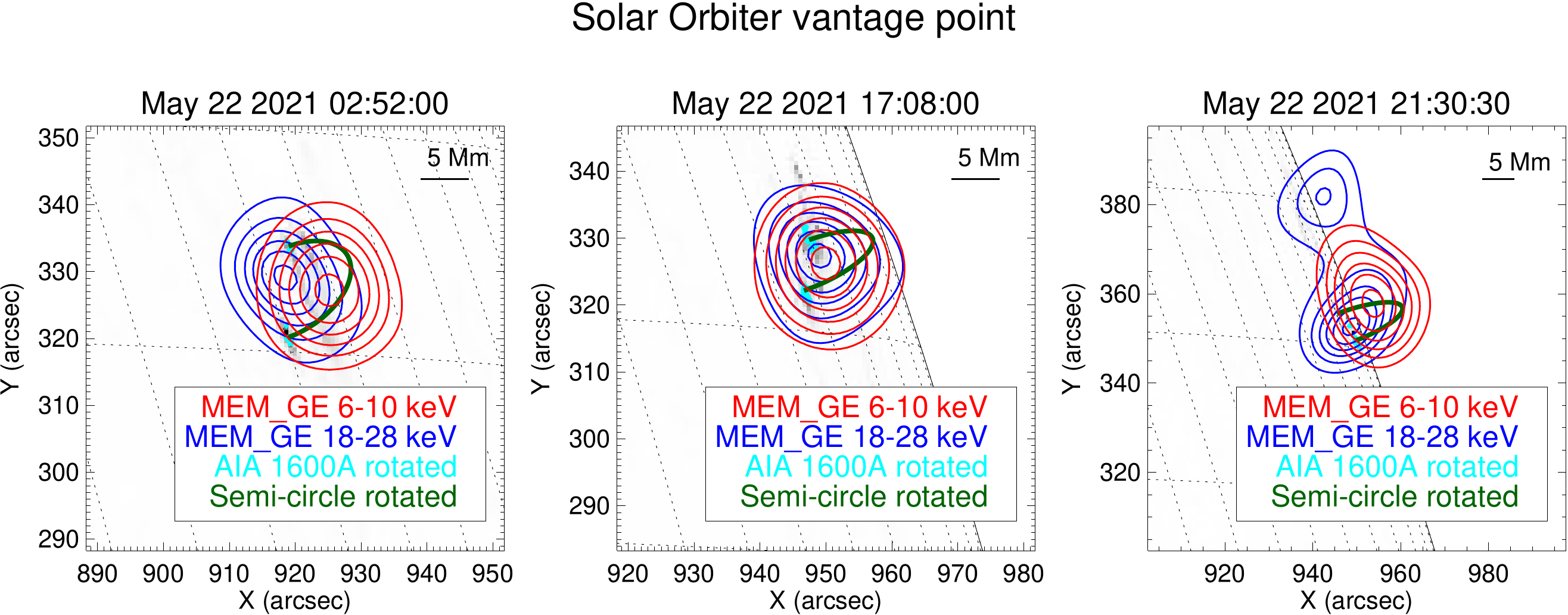}
\caption{Same as Figure \ref{fig:active-region-1} for the May 22 events recorded at 02:52:00 UT, 17:08:00 UT, and 21:30:30 UT.
The energy intervals considered are 6-10 keV for the thermal components and 18-28 keV for the non--thermal components.
}
\label{fig:active-region-2}
\end{figure*}

\begin{figure*}[!ht]
\centering
\includegraphics[height=\textwidth, angle=90]{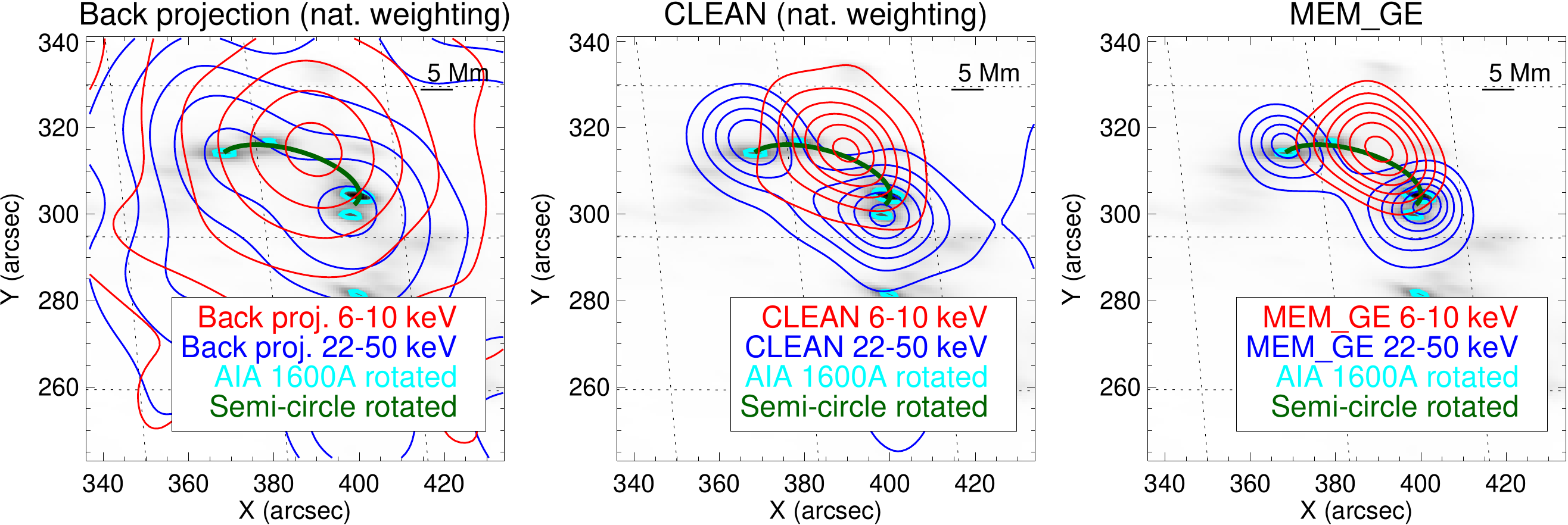}\\
\vspace{10pt}
\includegraphics[height=\textwidth, angle=90]{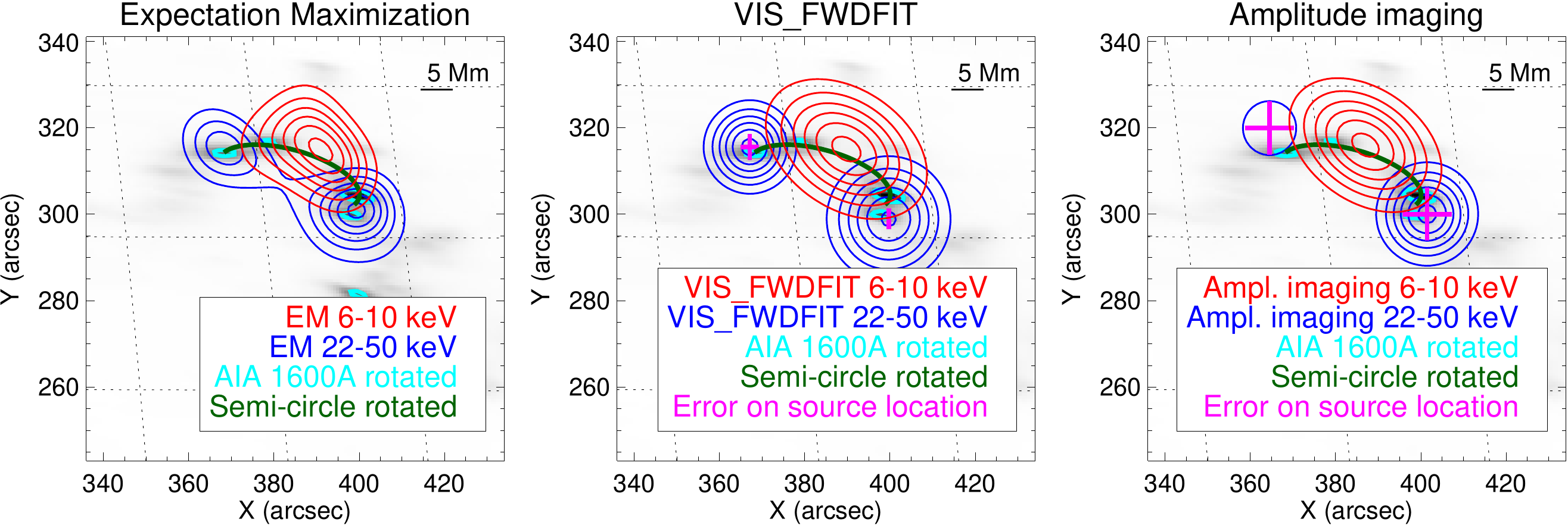}
\caption{From top to bottom, left to right: reconstructions of the May 7 2021 event provided by Back Projection, Clean, MEM$\_$GE, EM, VIS$\_$FWDFIT and the amplitude fitting method. The STIX reconstructions are overlaid on the rotated AIA 1600 \AA{} map of the same event.
Energy intervals, contour levels, semi--circles and bars indicating the 5 Mm length on the plane of the sky are the same as in Figure \ref{fig:active-region-1}.
In the VIS$\_$FWDFIT and amplitude fitting panels, the magenta crosses indicate the estimated error on each source location.
}
\label{fig:May-7-event}
\end{figure*}

Figure \ref{fig:active-region-1}, top row, contains the UV maps of the May 7, 8 and 9 events provided by AIA at 1600  \AA{}. In the bottom row of the same figure, the UV maps have been rotated in order to account for the Solar Orbiter vantage point and contour levels of the MEM$\_$GE reconstructions have been superimposed.
The same procedure has been followed in Figure \ref{fig:active-region-2}, in the case of the May 22 events.
In order to give an idea of possible flare loop geometry, a schematic vertical semi--circle connecting the ribbons is plotted in each panel of Figures \ref{fig:active-region-1} and \ref{fig:active-region-2}.

\begin{figure*}
\centering
\includegraphics[height=\textwidth, angle=90]{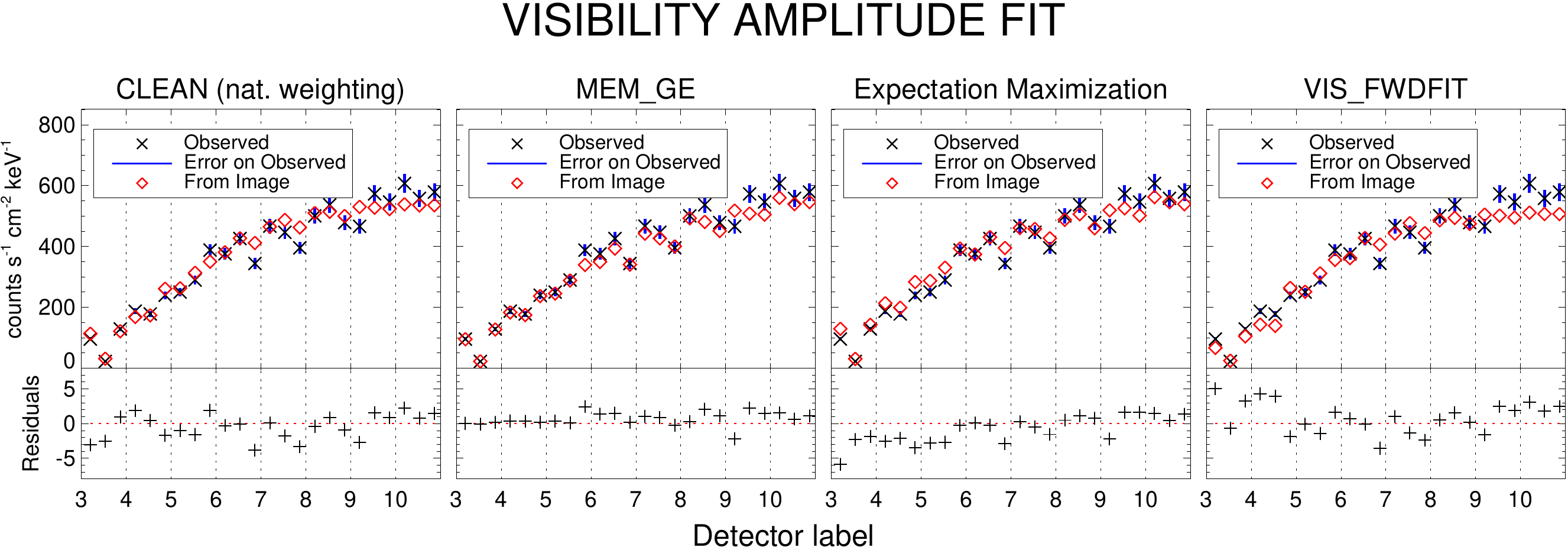}\\
\vspace{10pt}
\includegraphics[height=\textwidth, angle=90]{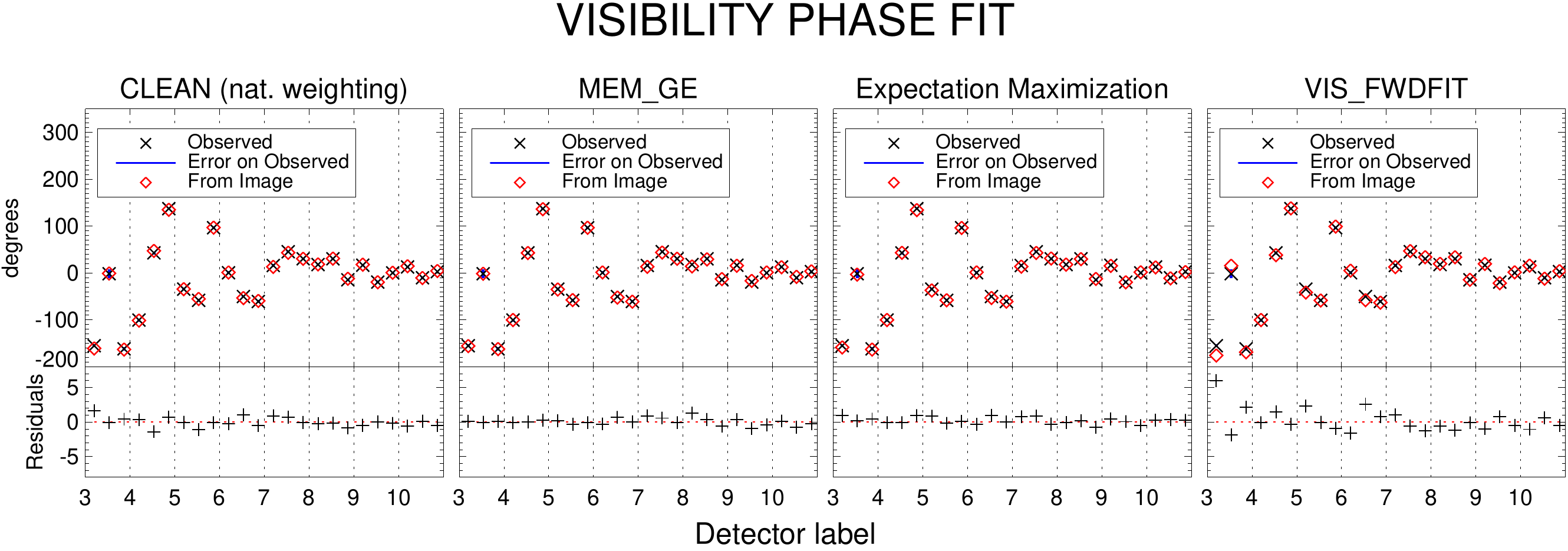}
\caption{Visibility amplitude fit (top panels) and visibility phase fit (bottom panels) of the thermal component reconstructions of the May 7 2021 event provided by Clean, MEM$\_$GE, EM and VIS$\_$FWDFIT (from left to right, respectively). The black crosses indicate the measured data, the blue bars represent the experimental uncertainty and the red diamonds denote the data predicted from each reconstruction.
The fit residuals are reported below each panel.}
\label{fig:fit-May-7-thermal}
\end{figure*}

\begin{figure*}
\centering
\includegraphics[height=\textwidth, angle=90]{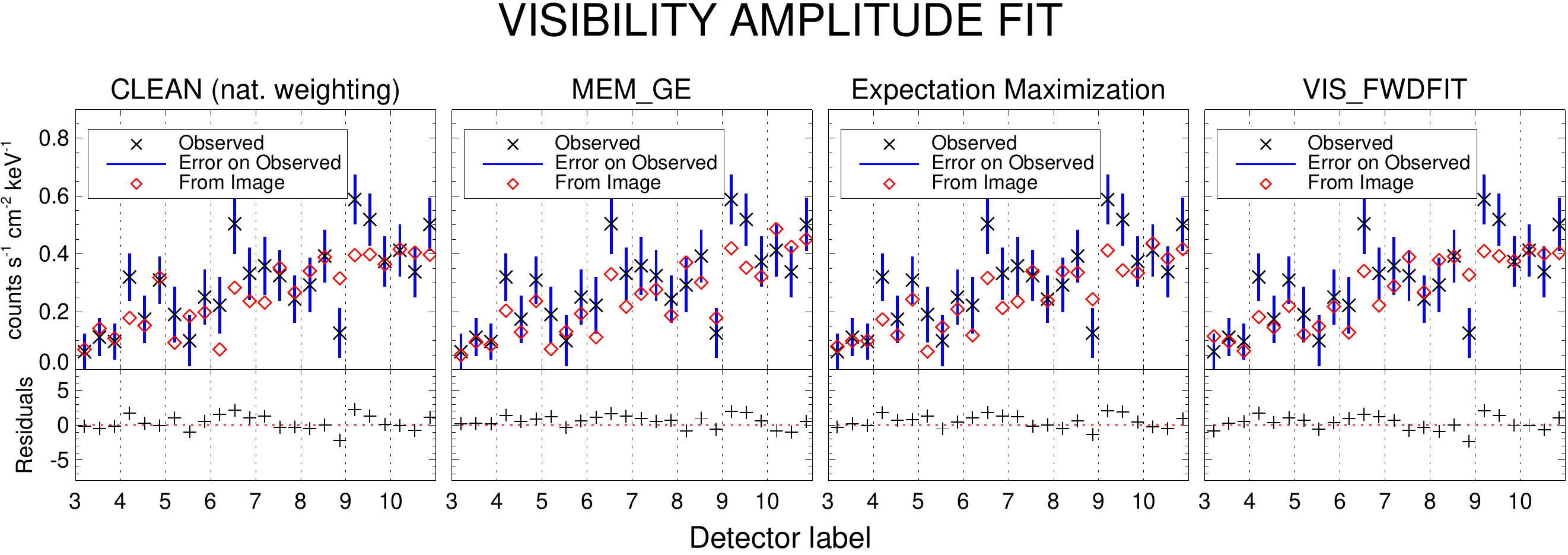}\\
\vspace{10pt}
\includegraphics[height=\textwidth, angle=90]{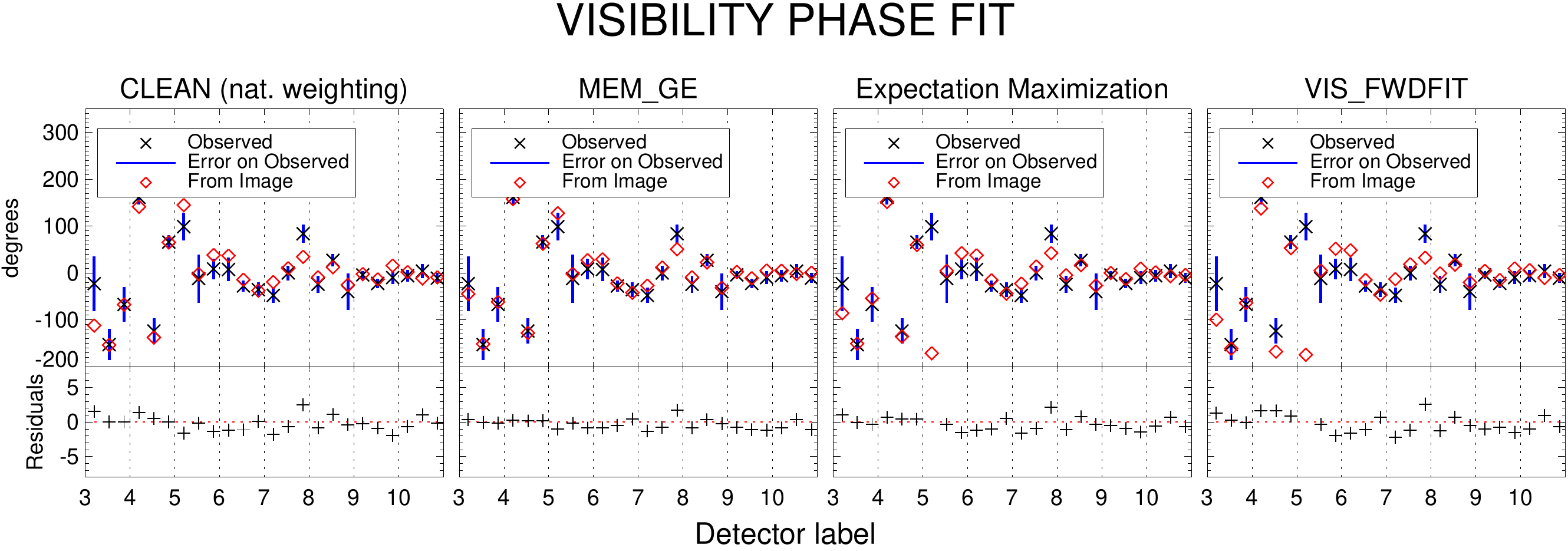}
\caption{Same as Figure \ref{fig:fit-May-7-thermal} for the non--thermal component of the May 7 2021 event.
}
\label{fig:fit-May-7-non-thermal}
\end{figure*}

Figure \ref{fig:May-7-event} compares the reconstructed images for the six methods in the case of the May 7 2021 event.
The STIX reconstructions have been superimposed on the rotated AIA 1600 \AA{} map.
As for the reconstructions of the non--thermal component, we report the uncertainty on the location of the reconstructed sources just for VIS\_FWDFIT and the amplitude fitting method.
Indeed, these are the only two methods that provide this information, since they rely on a parametric formulation of the image reconstruction problem. 
In Figures \ref{fig:fit-May-7-thermal} and \ref{fig:fit-May-7-non-thermal} we compared the amplitude and phases of the experimental visibilities with the ones predicted from the Clean, MEM$\_$GE, EM and VIS\_FWDFIT reconstructions of the thermal and non--thermal components.
In each panel, the data are ordered with respect to the corresponding detector label, whose number is reported in the abscissa. Detectors with the same resolution are ordered from left to right according to the label letter a, b or c.
As expected, the observed amplitude values shown in the plots are decreasing with respect to the detector resolution.
For instance, if we assume that the source has a Gaussian shape, then its bidimensional Fourier transform is still a Gaussian function.
Hence, the visibility amplitudes decay with a rate which is inversely proportional to the resolution of the sampling frequencies.
We also note that the observed visibility phases are close to zero for the coarsest sub--collimators, as is expected when the map center in the visibility calculation is selected to be at the flare location.
Below each plot we reported the residuals, i.e. the difference between the measured data and the ones predicted from the reconstruction, normalized by the errors.
The $\chi^2$ values of the reconstructions provided by the same algorithms are shown in Table \ref{tab:chi2-May-7}.
We note that, for Clean, the predicted visibility amplitudes and phases and the $\chi^2$ values are computed by using the derived Clean Components.
In this analysis we did not include the fits and the $\chi^2$ corresponding to the amplitude fitting method, because it does not consider visibility phases, and the ones corresponding to Back Projection, because it is a direct inversion without any regularization.

\begin{table}[ht]
\centering
\begin{tabular}{ccc}
\toprule
Method          &$\chi^2$ thermal   &$\chi^2$ non--thermal\\
\midrule
Clean           &3.95               &2.33\\
MEM$\_$GE       &1.73               &1.54\\
EM              &5.08               &2.04\\
VIS$\_$FWDFIT   &8.39               &2.84\\
\bottomrule
\end{tabular}
\caption{$\chi^2$ values associated to the reconstructions of the thermal and the non--thermal emission provided by Clean, MEM$\_$GE, EM and VIS$\_$FWDFIT on the May 7 2021 event.
The $\chi^2$ values are computed with respect to the observed visibilities.
We note that the value of the $\chi^2$ is of limited statistical relevance due to the not yet finalized error estimates.
Hence, they should only be compared between the different methods.}
\label{tab:chi2-May-7}
\end{table}

For a quantitative comparison of the results of the different methods, we reported in Table \ref{tab:param-ratio-fluxes} the flux ratio of the non--thermal footpoints reconstructed by the different methods.
The top--left footpoint is referred to as \emph{first source}, while the bottom--right footpoint is referred to as \emph{second source}.
As for the Clean algorithm, we used the Clean Components map for this analysis in order not to bias the results with the choice of the beam width used in the final convolution.

\begin{table}[h]
\centering
\begin{tabular}{ cc}
\toprule
Method           & Ratio fluxes \\
\midrule
Clean            & 0.574 \\
MEM\_GE          & 0.558 \\
EM               & 0.526 \\
VIS\_FWDFIT      & 0.560 \\ 
Amplitude fitting    & 0.448 \\
\bottomrule
\end{tabular}

\caption{Ratio between the flux of the first and the second source of the May 7 2021 event reconstructed by Clean, MEM\_GE, EM, VIS\_FWDFIT and by the amplitude fitting method. 
}
\label{tab:param-ratio-fluxes}
\end{table}

Finally, we show in Table \ref{tab:param-fwdfit-May-7} the values of the parameters related to the dimension, orientation and intensity of the sources reconstructed by the two forward--fitting algorithms, i.e. VIS\_FWDFIT and the amplitude fitting method.
We do not report the absolute location of the sources since it is not possible to retrieve this information from the visibility amplitudes only.
Indeed, the amplitude fitting method uses source configurations whose center is fixed in the origin of the coordinate system.

\begin{table*}[h]
\centering

\begin{tabular}{rcccc}
\toprule
\multicolumn{5}{c}{May 7 2021 - Thermal component}\\
\midrule
                    &\multicolumn{2}{c}{Amplitude fitting}                &\multicolumn{2}{c}{VIS$\_$FWDFIT} \\
\cmidrule{2-5}
Flux (counts s$^{-1}$ keV$^{-1}$ cm$^{-2}$)                &\multicolumn{2}{c}{$514.93 \pm 7.44$}    &\multicolumn{2}{c}{$516.05 \pm 8.41$} \\
FWHM max (arcsec)           &\multicolumn{2}{c}{$25.8 \pm 0.31$}      &\multicolumn{2}{c}{$27.1 \pm 0.48$} \\
FWHM min (arcsec)           &\multicolumn{2}{c}{$17.2 \pm 0.28$}      &\multicolumn{2}{c}{$18.1 \pm 0.32$} \\
Orientation angle (degrees)   &\multicolumn{2}{c}{$145.7 \pm 1.82$}     &\multicolumn{2}{c}{$148.9 \pm 2.16$} \\
\midrule
\midrule

\multicolumn{5}{c}{May 7 2021 - Non--thermal component}\\
\midrule
                    &\multicolumn{2}{c}{Amplitude fitting}                &\multicolumn{2}{c}{VIS$\_$FWDFIT} \\
                    &First source       &Second source      &First source       &Second source\\
\cmidrule{2-5}
Flux (counts s$^{-1}$ keV$^{-1}$ cm$^{-2}$)                &$0.14 \pm 0.07$    &$0.31 \pm 0.10$    &$0.15 \pm 0.03$    &$0.27 \pm 0.03$\\
FWHM (arcsec)                &$20.7 \pm 23.35$   &$15.7 \pm 6.68$    &$14.1 \pm 5.52$    &$19.2 \pm 3.66$\\
\bottomrule
\end{tabular}

\caption{Parameter values retrieved by the amplitude fitting method and by VIS$\_$FWDFIT on the May 7 2021 event. Top: parameters values of a fitted elliptical Gaussian shape. Bottom: parameters values of a fitted double circular Gaussian shape.}
\label{tab:param-fwdfit-May-7}
\end{table*}

\subsection{Discussion of results}

The X-ray emissions reconstructed from STIX data are consistent with the ribbons shown in the AIA maps in terms of separation, shape and orientation (see Figures \ref{fig:active-region-1} and \ref{fig:active-region-2}).
In particular, the reconstructions of the May 7 2021, May 8 2021 and May 9 2021 events are compatible with a standard flaring configuration with two chromospheric footpoints and a coronal loop--top source.
For these events, the relative position between the thermal and the non--thermal emission is in agreement with the height and the direction of the semi--circles connecting the ribbons, at least in projection. 

The reconstruction of the first two flares on May 22 reveals a rather compact source size, and the non--thermal sources appear unresolved. 
Once fully calibrated, the finest sub--collimators should be used to potentially separate the flare ribbons in the non--thermal image. 
While the first May 22 flare clearly shows a separation of the thermal and non--thermal emission, the event around 17UT shows both emissions originating from close to the same location. Co--spatial thermal and non--thermal sources are occasionally observed in so--called thick target coronal sources (Veronig \& Brown 2004). 
Coronal thick target sources tend to have soft (steep) non--thermal spectra, which is similar to what is observed in this flare (see Figure 2). In the case that this event would indeed be a coronal thick target, the X-ray sources would come from the corona, and our aligment by eye would need to be adpated accordingly. Further analysis is in progress investigating this flare. 

For the May 22 flare around 21UT, the brighter of the two non--thermal sources appears to be a superposition of the two main sources from the two flare ribbons seen in projection. The weaker non--thermal source to the north (Figure 4, bottom right) might be a secondary non--thermal source on the extension of the eastern flare ribbon towards north. This clearly shows the value of combining observations of different look directions to fully understand the flare geometry. 


The five imaging methods from calibrated visibilities provide consistent results on the May 7 2021 event, particularly respect to the location of the reconstructed sources and the ratio between the footpoint fluxes (see Figure \ref{fig:May-7-event} and Table \ref{tab:param-ratio-fluxes}).
The MEM$\_$GE, EM and VIS$\_$FWDFIT reconstructions of the thermal emission present very similar dimension and orientation and those of the non--thermal emission show two footpoints with comparable size, orientation and separation.
The Back Projection reconstructions, as expected, present sidelobes due to the limited number of Fourier components sampled by STIX and to the fact that this algorithm directly inverts the data, without applying any regularization \citep{2002SoPh}.
Clean provides reconstructions with much reduced noise levels with respect to Back Projection, however the dimensions of the retrieved sources can be larger when compared to the results of MEM$\_$GE, EM and VIS$\_$FWDFIT. This behavior is intrinsic to the Clean algorithm because the derived point source model (the Clean Components) are convolved with an idealized PSF (the Clean beam).
There are various choices for the selection of the beam size.
The most conservative case is to approximate the core of the PSF of input image (dirty map) with a Gaussian function. 
This choice gives a beam size of 31.3$''$ FWHM, which would smear out the Clean maps significantly. 
In the other extreme, the resolution provided by the finest sub--collimators used in the image reconstruction could be used to approximate the clean beam size. For the Clean images presented here this would give a beam size of 14.6$''$ FWHM. For such compact beam sizes, Clean boxes around the potential solar sources should be used to avoid having noise peaks wrongly identified as solar sources. If by mistake a noise peak is cleaned and convolved with such a narrow beam, noise features can be significantly enhanced compared to Back Projection map, making the resulting clean image look questionable. For this paper we used a beam size of 16.5$''$ FWHM, using Clean boxes derived from the 50\% contours. The use of a smaller Clean beam could result in a better match of the Clean sources size compared to the other imaging algorithms. 
While the output of Clean (i.e. the list of Clean Components) is independent of the selected beam size, the visualization of the Clean result (i.e. the Clean image) is affected by the choice of the clean beam. The selection of the Clean beam width should be adopted depending on the size of the sources within the image. For compact footpoint sources, a small beam size is a good option, while for extended sources, a small beam size can artificially break up the source, and it is better to use a larger beam size.

From the data fitting shown in Figures \ref{fig:fit-May-7-thermal} and \ref{fig:fit-May-7-non-thermal} and from the low $\chi^2$ values shown in Table \ref{tab:chi2-May-7}, we deduce that the algorithms from calibrated data are able to fit the experimental visibilities with high accuracy.
MEM$\_$GE systematically outperforms the other methods in terms of data--fidelity (see Table \ref{tab:chi2-May-7}).
The good performance of this algorithm is also due to an ad hoc choice of the regularization parameter, whose values are shown in Table \ref{tab:param-algo}. Changing parameters in the other algorithms, such as the selection of Clean boxes, can be used to optimize the image quality. 
Clean and EM provide reconstructions of the non--thermal emission with a comparable $\chi^2$ value, while the visibility--based method has a slightly better performance with respect to the count--based one in the reconstruction of the thermal emission.
Finally, the $\chi^2$ values associated with the VIS$\_$FWDFIT reconstructions are consistently the highest ones.
This is most likely due to the fact that the shapes used for fitting do not always represent the true morphology accurately.

Finally, as far as the comparison between the two forward--fitting algorithms (VIS\_FWDFIT and the amplitude fitting method) is concerned, Table \ref{tab:param-fwdfit-May-7} shows that the parameters retrieved by the two methods are very similar in the case of the thermal emission.
However, the source reconstructed by VIS\_FWDFIT is slightly larger and tilted with respect to the one reconstructed by the amplitude fitting method.
Instead, the non--thermal footpoints reconstructed by the two methods have comparable fluxes but different FWHM; nevertheless, the discrepancy between the retrieved FWHM values is compatible with the associated uncertainty, which is particularly large in the case of the first source for the amplitude fitting method.
This large uncertainty is possibly due to the fact that amplitude fitting utilizes half of the data with respect to VIS\_FWDFIT consisting of just the visibility amplitudes; consequently, it suffers from a more pronounced numerical instability.
 
\section{Conclusions}

We presented the first results of STIX imaging using the best current calibration and imaging software implementation as of January 2022. 
Specifically, we showed that the visibility phases of the 24 coarsest detectors are well-calibrated since images reconstructed from STIX visibilities are reliable in terms of morphology, dimension and orientation when compared to the flare ribbons seen in the AIA 1600 \AA{} maps of the same events.
We compared the performances of several algorithms implemented for the solution of the STIX imaging problem from calibrated data, showing consistent results and a good accuracy in reproducing the experimental visibilities.
Finally, we provided a further validation of the phase calibration by showing good agreement between the reconstructions obtained from visibility amplitudes and those obtained from calibrated data.
Future work will be devoted to further improve the existing calibration, as well as a first calibration of the data recorded by the finest six sub--collimators at 7 and 10 arcsec resolution. 
We point out that the implementation of other reconstruction methods is under construction and involves a Sequential Monte Carlo scheme \citep{sciacchitano2018identification, sciacchitano2019sparse}, compressed--sensing methods \citep{duval2018solar,felix2017compressed} and another EM-like approach for counts \citep{2020OAst...29..220S}.

\begin{acknowledgements}
{\em{Solar Orbiter}} is a space mission of international collaboration between ESA and NASA, operated by ESA. The STIX instrument is an international collaboration between Switzerland, Poland, France, Czech Republic, Germany, Austria, Ireland, and Italy. AFB, HC, GH, MK, HX, DFR and SK are supported by the Swiss National Science Foundation Grant 200021L\_189180 and the grant 'Activités Nationales
Complémentaires dans le domaine spatial' REF-1131-61001 for STIX. PM, EP, FB and MP acknowledge the financial contribution from the agreement ASI-INAF n.2018-16-HH.0. 
SG acknowledges the financial support from the "Accordo ASI/INAF Solar Orbiter: Supporto scientifico per la realizzazione degli strumenti Metis, SWA/DPU e STIX nelle Fasi D-E".
\end{acknowledgements}

\bibliographystyle{aa.bst}
\bibliography{bib_stix}

\end{document}